%%%%%%%%%%%%%%%%%%%%%%%%%%%%%%%%%%%%%%%%%%%%%%%%%%%%%%%%%%%%%%%%%%%%%%%%%%%%%%
% This is a TEX file. 10 pages. Uses epsf, harvmac. 1 postscript figure.
% Authors: Indranil Dasgupta and Ryan Rohm.
% Preprint number: BUHEP-96-39
% Revised on: September 26, 1996.
%%%%%%%%%%%%%%%%%%%%%%%%%%%%%%%%%%%%%%%%%%%%%%%%%%%%%%%%%%%%%%%%%%%%%%%%%%%%%%

\input harvmac.tex
\input epsf.tex
\overfullrule=0pt
% defs.tex
\def\caption#1{{\it
        \centerline{\vbox{\baselineskip=12pt
        \vskip.15in\hsize=4.2in\noindent{#1}\vskip.1in }}}}

 1
 1
 1

% titleII.tex
\lref\rohm {
R. Rohm, I. Dasgupta, Phys. Rev. {\bf D 53}, 1827 (1996)}

\lref\branden {
R. H. Brandenberger, A. Davis, M. Hindmarsh, Phys. Lett. {\bf B 263 },
239 (1991).}

\lref\Polyakov{
A. M. Polyakov, Nucl. Phys. {\bf B 268}, 406 (1986).}

\lref\turok{
T. W. B. Kibble, N. Turok, Phys. Lett. {\bf B 116}, 141 (1982).}

\lref\action {
D. Forster, Nucl. Phys. {\bf B 81}, 84 (1974);
K. Maeda, N. Turok, Phys. Lett. {\bf B 202}, 376 (1988);
R. Gregory, Phys. Lett. {\bf B 206},  199 (1988);
P. Orland, Nucl. Phys. {\bf B 428}, 221 (1994).}

\lref\vach{
D. Garfinkle, T. Vachaspati, Phys. Rev. {\bf D 36}, 2229 (1987).}

\lref\Vilenkin{
A. Vilenkin, Phys. Reports {\bf 121, 5}, 263 (1985).}

\lref\Orland{
P. Orland, Nucl. Phys. {\bf B 428}, 221 (1994).}

\lref\grav{
J. M. Quashnock, D. N. Spergel, Phys. Rev. {\bf D 42},
2505 (1990).}

\lref\ruth {
R. Gregory, hep-ph/ 9412232.}

\lref\others{
T. Vachaspati, A. Vilenkin, Phys. Rev. {\bf D 31}, 3052 (1985);
T. Vachaspati, A. Vilenkin, {\it ibid}. {\bf 35}, 138 (1987).}

\lref\moreref {
R. H. Brandenberger, Nucl. Phys. {\bf B 293}, 812 (1987);
P .Bhattacharjee, Phys. Rev. {\bf D 40}, 3968 (1989);
R. H. Brandenberger, A. Sornborger, M. Trodden, Phys. Rev.  {\bf D 48},
940 (1993).}

\lref\numer {
E. P. S. Shellard, in {\it {Cosmic Strings - The Current Status}},
Proceedings of the Yale Cosmic String Workshop, New Haven, Connecticut,
1988. Edited by F.S.Accetta and L.M.Krauss. (World Scientific,
Singapore, 1989) p.25.
}

\Title{\vbox{\baselineskip12pt\hbox{BUHEP-96-39}\hbox{hep-ph/9610201}}}
{\vbox{\centerline{Binary String Dynamics}\vskip 2pt}}
\vskip .5in
{\parskip=0pt\baselineskip=12pt
\centerline{ Indranil Dasgupta$^{1,a}$ {\it {and}} Ryan Rohm$^{1,b}$ }\footnote{} {$^1$\it e-mail:
dgupta@budoe.bu.edu, rohm@ryan.bu.edu}
\bigskip
\centerline{\sl $^a$ Boston University, Boston,
Ma 02215, USA}
\centerline {\sl $^b$ University of North 
Carolina, Chapel Hill, NC  27599, USA}}
\bigskip

\vskip .5in

{In this paper we investigate the
dynamical properties of binary cosmic strings \rohm. 
We find extrinsic curvature dependence of the string action and show
that kinks on binary strinks are eroded while cusps can play a
major role in their evolution.}

\Date{September 26, 1996}
%\draftmode

%\draft

% intro.tex
\newsec{Introduction}

In a recent paper \rohm$\,$
 it was shown that the cosmic string solutions
of some grand-unified theories have a microscopic structure which is not
cylindrically symmetric, but instead is flattened, giving rise to a
preferred transverse direction.  In this paper we investigate the
dynamical properties of these binary cosmic strings. 
We begin by deriving an effective action for the binary string, using an
approximation whereby two worldsheets defined by the cores of the two
string components are taken to have constant (spacelike) separation.
Describing the string components by the Nambu-Goto action,
we then express the result in terms of one worldsheet, with
a transverse spacelike vector determining the separation of the
component strings. The classical equations obtained in this description 
display a
macroscopic motion close to that of a single string, while the
separation vector leads to an extra light degree of freedom for
the string, with short-range interactions.  The equations are shown to
admit a natural  geometrical description in terms of the extrinsic
geometry of an embedded surface. Because of the difference in the 
relevant length scales, one may,
to first approximation, consider the bulk motion of the binary string
to be independent of the local separation vector and consider the
dynamics of the separation vector as it evolves on a background 
geometry defined by a solution to the classical string equations.
In this vein it is also possible to consider the effects of 
kinks, cusps, {\it etc.}

% revsptII.tex

\newsec{Extrinsic Curvature Dependence of the Action.}

The effective action of long cosmic 
strings can be obtained by integrating out all
high energy modes in a background containing a string-like defect
\Polyakov , \Orland. The resulting action has only the collective
coordinates corresponding to the space-time positions of the string. The
simplest approximations give the well-known Nambu-Goto action \action 

\eqn\wldsh{
S= T{\int}d\sigma_1 d\sigma_2 {\sqrt {-det D}}\, , 
}
where T is the string tension (the mass per unit length of the string), 
$D_{ij}={\partial x^\mu\over \partial \sigma_i}{\partial
x_\mu \over \partial \sigma_j}$ ; $i,j =1,2$. A {\it binary}
string consists of two cosmic strings bound together by a potential (that
may be confining) \rohm . Its action can be written as 
the sum of the individual Nambu-Goto actions of
the constituent strings plus an interaction term, i.e, $S= S_1 + S_2 -
V$. 
If the interaction $V$ is sufficiently strong to force
the `spacing' of the strings to be constant, the effective
action may be reduced to the geometric quantity $S_1+S_2$ with the
additional constraint on the `spacing' of the component strings.

We will first give a more precise meaning to the term `separation'.
When the two world sheets are
{\it sufficiently flat and parallel}$\,^2$\footnote{}
{$^2$ A more complete formulation over the bulk of the string motion is 
irrelevant, and morever relatively intractable.
Local behaviour of the world sheets is well described by 
small deformations of the flat and parallel world sheets.}
 we can draw 
at each point $p$ on the world sheet $\Sigma _1$  a space-like normal that
intersects the world sheet $\Sigma _2$ once.
 This induces a map from $\Sigma_1$ to $\Sigma_2$ which allows us to
parametrize
$\Sigma_1$ and $\Sigma _2$ by the same coordinates $\sigma _1$ and $\sigma _2$.
Then locally at least,
\eqn\loccoord{
x^{\mu}_{\Sigma_2}(\sigma _1,\sigma _2)=x^{\mu}_{\Sigma _1}(\sigma _1,\sigma 
_2)+a^{\mu}(\sigma
_1,\sigma _2)\, , 
}
where the `separation' vector 
$a^{\mu}$ is a section of the normal bundle on $\Sigma_1$ at ($
\sigma_1, \sigma_2 $). 
The map exists
locally for an interesting region of 
small perturbations around flat and parallel worldsheets and may admit
multiple realizations.

With this parametrization the sum of the areas of the two
string world sheets can be effectively reduced to an integral over a
single world sheet. However, superior results are obtained by 
using a world sheet lying `midway' between  $\Sigma_1$
and  $\Sigma_2$ as the basis. 
We shall refer to this as the effective binary string world
sheet $\Sigma$. Choosing coordinates $\sigma_1$ and $\sigma_2$ on $\Sigma$ one
can parametrize $\Sigma_1$ and  $\Sigma_2$ by the same coordinates using
{\it two} normal
vectors $a_1^{\mu}$ and $a_2^{\mu}$ on $\Sigma$. Locally we can define $\Sigma$
such that the `center of mass' condition $T_1a_1^{\mu}+T_2a_2^{\mu}=0$
is valid. Clearly, the `spacing' between the string
components can be fixed by
constraining the magnitudes of the vectors $a_1$ and $a_2$ to be constant.
Then the geometric action $S_1 + S_2$ reduces to,
 $S=\sum T_I M_I$, with $I=1,2$ and 
\eqn\Seff{
M_I = \int d\sigma_1 d\sigma_2 
\sqrt {-{\rm {det}}\left [ \left \{{\partial(x + a_I)^{\mu}\over 
\partial \sigma_i}\right \}
\left \{{\partial (x+a_{I})_{\mu} \over \partial \sigma _j}\right \} 
\right ] \, .
}}

When fluctuations of $a_I^{\mu}$ are small compared to the bulk motion of the 
binary string, the square roots can be expanded in powers of the
derivatives of $a_I^{\mu}$.
To develop the formalism we shall choose a zweibein $e_a^{\mu}$
($a=1,2$) and a basis for the normal space $n_k^{\mu}$ ($k=3,4$) on the
world sheet $\Sigma $. This tetrad satisfies the relations

\eqn\edotn{
\eqalign{
e_a^{\mu}e_{b\mu}= &\eta _{ab}\, ,  \cr
n_k^{\mu}n_{l\mu}=&-\delta _{kl}\, ,  \cr
e_a^{\mu}n_{k\mu}=&0 \, .\cr
}
}
Note that $\eta _{ab}$ has Minkowski signature. We will always
follow the convention of raising and lowering indices in the tangent
bundle by $\eta_{ab}$ and in the normal bundle by $-\delta_{ij}$. 
We choose the zweibein by defining, 
$ \partial_a x^{\mu}= e^{\mu}_a \sqrt{\partial_b x^{\nu} \partial_c
x_{\nu} \eta^{bc}}$. 
The affine connection $\Gamma_{ab}^c$ and the extrinsic curvature tensor
(or the second fundamental form) $K_{ab}^k$ are then defined by,
${
\partial_ae_b^\mu = \Gamma_{ab}^ce_c^{\mu}+K_{ab}^kn_k^{\mu}
}$. 

Expanding $a_I^{\mu}$ as $a_I^{\mu}=a_I^{k}n_k^{\mu}$ one finds that

\eqn\nderiv{
\partial_bn_k^{\mu}=K_{bk}^ae^{\mu}_a + A_{bk}^in_i^{\mu} \, , 
}
where $K_{bk}^a=-K^l_{bc} \delta _{kl}\eta ^{ca}$,
and $ A_{bk}^i$ is a connection on the normal bundle of the world
sheet. The curvature $F^l_{abk}$ of the connection $A^l_{bk}$ is given
by, $F_{abk}^l = K_{ak}^c K_{bc}^l - K_{bk}^c K_{ac}^l$, and is
generally non-zero. A similar statement holds for the curvature of the
connection $\Gamma_{ab}^c$.
The connections  $A^l_{bk}$ and  $\Gamma_{ab}^c$ can not 
therefore be locally gauged away.
Using the connection  $A^l_{bk}$
 we can define a covariant derivative $D^i_{kb}a_I^{k}$ as

\eqn\covdiff{
D^i_{kb}a_I^{k} = -D_{ikb}a_I^k = \partial_ba_I^{k}\delta ^i_k -
A_{bk}^ia_I^{k} \, .
}
To expand \Seff, we'll choose the gauge, $\sigma_1=t,
\sigma_2= \xi$, where $t$ is time and 
$\xi$ is the length of the string measured from a
particular point on the string.
Using $T_1a_1^{\mu}+T_2a_2^{\mu}=0$,
to relate $a_1$ and $a_2$ and 
defining $T= T_1 + T_2$, and $a^k =
a^k_1 \sqrt {(T_1/T_2) }$ we get

\eqn\newexp{
\eqalign{
S=\ & T{\int}dt d\xi (1-v_t^2)\left [1 - {1 \over 2}\left \{ {1 \over 1-v_t^2}
(D^i_{kt}a^k)^2 - (D^i_{k\xi }a^k)^2 +
(a^kJ^a_{ak})^2 - 2 det (a^k J^a_{bk})\right \}\right ] \cr
+\ &{\rm{higher \, order\, .}}\cr
}
}
Here $v_t$ is the transverse speed of the
string and $J^a_{bk}$ is defined by $J^{\xi}_{\xi k}=K^{\xi}_{\xi k},
J^{\xi}_{tk}=K^{\xi}_{tk}, J^t_{tk}\sqrt{1-v_t^2}=K^t_{tk},
J^t_{\xi k}\sqrt{1-v_t^2}= K^t_{\xi k}$.
The world sheet indices
$a$ and $b$ are contracted using the tensor $\eta ^{cd}$ while the
normal indices $k$ and $l$ are contracted using
$-\delta ^{mn}$. The determinant is taken over the matrix defined by the
indices $a$ and $b$.
Note that the only physical velocity is $v_t$ corresponding to the 
motion of the string perpendicular to itself.

The extrinsic curvature dependence is explicit. The
fields $a^k$ live on the world sheet and couple to the extrinsic
curvatures. The small parameters of
the expansion are the derivatives of $a^k$ {\it and} the eigenvalues of the
matrices $a^kK^a_{bk}$. 
The phenomenology  resembles
that of a thin band with the transverse vector $a$ identified as 
the `width' of the
band. The coupling to the extrinsic curvatures 
defines the elastic properties of the band
associated with bending of the band in its own plane and 
in a normal plane. Of particular interest are the twists of the band
which now constitute a new propagating mode on the string.

\newsec {Twist Modes }

The classical equation of motion
for the (unadorned) cosmic string with the 
empirical Nambu-Goto action can be described as follows \turok, 
\Vilenkin. The world sheet is described in 
Minkowski space by $x^{\mu}(\xi,t)$ with $\xi $ and $t$ being 
world sheet coordinates. This has the usual reparametrisation invariance
$\xi, t \rightarrow \xi ^ {\prime} ,  t ^{\prime}$.
Choosing the gauge conditions: $ \, t = x^0, \, \left({\partial  x^i \over 
\partial  t }\right )\left({\partial  x^i \over
  \partial  \xi }\right ) = 0, \, \left({\partial  x^i \over \partial  t
}\right )^2 - \left({\partial
  x^i \over \partial  \xi  }\right )^2 = 1$,
one obtains the following equation of motion for the spacelike
components of $x^{\mu}(\xi,t)$:

\eqn\xeom{
{\partial ^2 x^i \over \partial  t ^2 } - {\partial ^2 x^i  \over 
\partial  \xi ^2 } = 0 \, .
}
The solution to \xeom$\,$ is

\eqn\solconstr{
x^i(\xi ,t) = 1/2 [ c^i (\xi - t) + d^i(\xi + t)] 
}
with, 
$\left ({\partial  c^i \over \partial  \xi }\right )^2 = 
\left ({ \partial  d^i \over \partial  \xi }\right )^2 = 1
$. Here ${\partial  c^i \over \partial  \xi} $ and ${ \partial  d^i \over
\partial  \xi }$ are components of 
spacelike three vectors drawn from the origin with
tips lying on the Kibble-Turok sphere of unit radius \turok .

In the present case, we assume the bulk motion of the binary string
to be unaffected by its binary character and look at the
equation of motion of $a$ with a given background solution for the bulk
motion of the string. 
Substituting ${{\partial  x^ \mu  }\over \partial  \sigma
  _a }/ {| {\partial  x^ \mu  \over \partial  \sigma _a} |}$
for $e^{\mu} _ a$ in the 
definition of the second fundamental form  (with $ \sigma _a = \xi , t)$ 
we have, 
$ K_{ab}^l = -(\partial _a e_b^{\mu })n_{k \mu } \delta ^{kl}=
{-\partial _a (\partial _b x^{\mu })n_{k \mu } \delta ^{kl}
 \over {| \partial_b x^{\mu }|}}$.
In the gauge we have chosen, 
$ |\partial _{\xi }x^{\mu }| = |\partial _tx^{\mu }|$.
This implies that $ K_{\xi t}^l = K_{t \xi }^l =
\beta ^l$. Using \xeom$\,$ we also get $K_{\xi \xi}^l = K_{tt}^l =
\alpha ^l$. With this form of $K^a_{bk}$, we obtain the following equation 
of motion for $a^{k}$:
\eqn\aeom{
{1 \over 1-v_t^2}(D^i_{kt}D_{ijt}a^j) -
(D^i_{k \xi}D_{ij \xi} a^j) -
2(a^i \alpha_i)\alpha_kQ_1 +2 (a^i \beta_i)\beta_kQ_2= 0 \, , 
}
where $Q_1 = \left [{1 \over \sqrt {1-v_t^2}} -{1 \over 2} (1+{1 \over \sqrt
{1-v_t^2}})\right ]$ and $Q_2 = \left [{1 \over \sqrt {1-v_t^2}}\right ]$.
Terms involving $\alpha_k$ and
$\beta_k$ can be dropped  when looking for traveling
wave solutions of large momentum. 
One then obtains a free wave equation implying
that the coupling to extrinsic curvatures is not relevant for small
wavelength excitations.
For wavelengths comparable to the scale
of the string curvatures, the extrinsic curvatures of the world sheet 
couple strongly to the $a^k$. {\it  Thus energy from the bulk motion of the
string can be transferred to long wavelength winding or twisting
motions of $a^k$ }. 
Note that the interaction between the two components of the
binary string is taken care of by constraining the magnitude of $a$ to
be constant. This still 
leaves an angular coordinate (describing the `twist' of the binary
string) free.

The phenomenology of the twist modes can be explored with particular
background bulk motions of the string.
As an example consider the closed string solution \Vilenkin
\eqn\strclps{
{\bf x} =  L [ {\rm {sin}}(\phi ){\rm {cos}}(\rho ) \hat {e_1} + 
                      {\rm {cos}} (\phi ) {\rm {cos}}(\rho ) \hat {e_2}]
\, , 
}
where, $ L =  2 \pi \times $mass of the string, $ \phi = { \xi \over L }$,
$ \rho = { t \over L }$ and $\hat {e_1}$ and $\hat {e_2}$ are
orthonormal vectors.
The string looks like a collapsing or expanding circle with
maximum circumference equaling $ { L \over \mu } $, where $ \mu =
T_1 + T_2 $, is the mass per unit length.

For this family of solutions $\beta ^l=0$, 
$A=0$. Also $v_t^2 = {\rm {sin}}^2(t/L)$. 
We define the twist mode $\theta$ to be the
angle between $a$ and the $z$ axis.
The second fundamental form $K^l_{ab}$ has only one
non-trivial element, $\alpha^n = -n.{\partial ^2_t x \over |\partial_t x
|}= (1/L) {\rm {sec}}(t/L)$
where $n$ is the unit normal to the world sheet and to the $z$ axis.
Varying only the
angular coordinate $\theta$ in action \newexp$\,$ we get the 
equation of motion
\eqn\thetamotion {
s^2 \partial_t ^2 \theta - \partial_{\xi}^2 \theta + 2 {\rm {sin}}
 \theta {s^2 \over L^2} [s - (1+s)^2] \, , 
}
where $s = {\rm {sec}}(t/L)$. The explicit time dependence in {\bf {$s$}}
causes energy exchange between bulk motion and winding oscillations. The
energy exchange is large and is likely to cause some kind of 
equipartition of energy between curvature oscillations and twist
modes. The effect on kinks, as we show now, is even more dramatic.

\newsec {The Fate of Kinks}

Kinks are points on the string where the diagonal element ($K^k_{\xi
\xi}$) of the second fundamental form is singular. From \solconstr$\,$
it is clear that kinks arise
from discontinuities in ${ \partial c^i \over \partial  \xi }$ or ${
\partial  d^i \over \partial  \xi }$.  Isolated kinks
travel to the `left' or `right' at the speed of
light.
It is therefore easier to write the kink solution in the light-cone
coordinates, $ u= \xi + t , v= \xi - t$.
The kink solution is proportional to $\delta (u)$ or $\delta (v)$
for a kink travelling in the backward $\xi$ or forward $\xi$ direction.
Precisely because of the singularity in the extrinsic curvature at the
kinks, the approximation used for deriving the twist mode equations
breaks down. However, physically, 
the thickness of a strongly bound binary string is $\sim \eta
^{-1}$, where $\eta ^{2}$ is the mass per unit length of the string and the
extrinsic curvature at the kinks is less than $\eta$.
So for the physical kinks, our scheme of approximation
remains valid.

When a long wavelength twist mode interacts with a kink we can
keep the $\delta$-function approximation for the kink. 
The transverse speed $v_t$ varies slowly along the string
and may be neglected ($ 1-v_t^2 \sim
1$). 
The equation of motion for $\theta$ then simplifies to
\eqn\thtltcon{
2{ \partial ^2 \theta  \over \partial u  \partial v } + 6{ \gamma \eta ^2 } 
\delta (v\eta )\rm {sin}(2\theta ) = 0 \, ,
}
where $\gamma $ is proportional to the angle of the kink.
To first order in $\theta $ the equation is linear and the solution is a
free wave (with frequency $\omega$) 
moving opposite to that of the kink and suffering a phase
shift after collision with the kink:
\eqn\knksol{
\theta _0 = A {\rm {exp}} (i \omega u +i{6 \gamma
{\eta \over \omega}}\Theta ({v \eta})) \, , 
}
where $\Theta ({v \eta})$ is the step function. 
To the next order the anharmonic term of order $A^3$ in \thtltcon$\,$
produces scattering. 
For large amplitudes the
small $\theta$ approximation fails, but
the total power dissipation is easily estimated
by noting that $\eta$ is the only mass
scale relevant to the problem. Hence the total power dissipation from a
kink is $\sim \eta ^2$. The energy in the kink is $\sim \eta$; so the
time taken for the kink to disappear is $\sim \eta ^{-1}$. This is a
very small time scale compared to the expansion times of large string
loops ($\sim L$).

We have solved \thtltcon$\,$ numerically for a straight cosmic
string with periodic boundary conditions. With a kink angle of 1
radian and a kink size of $4 \eta ^{-1}$ we simulated the scattering of
a sinusoidal twist wave of wavelength $4 \eta ^{-1} $ and amplitude $2
\pi$ traveling in a 
direction opposite to that of
the kink. In Figure 1 we plot the ratio of the energy lost by the kink to
the energy incident on it as a function of time. The ratio rises
to about 25 in the time scale of $4\eta ^{-1}$ and oscillates about that 
value, implying a quick draining of
energy from the kink. 

\topinsert
\vskip -1.0cm
{\centerline{\epsfxsize=1.5 in\epsfbox{
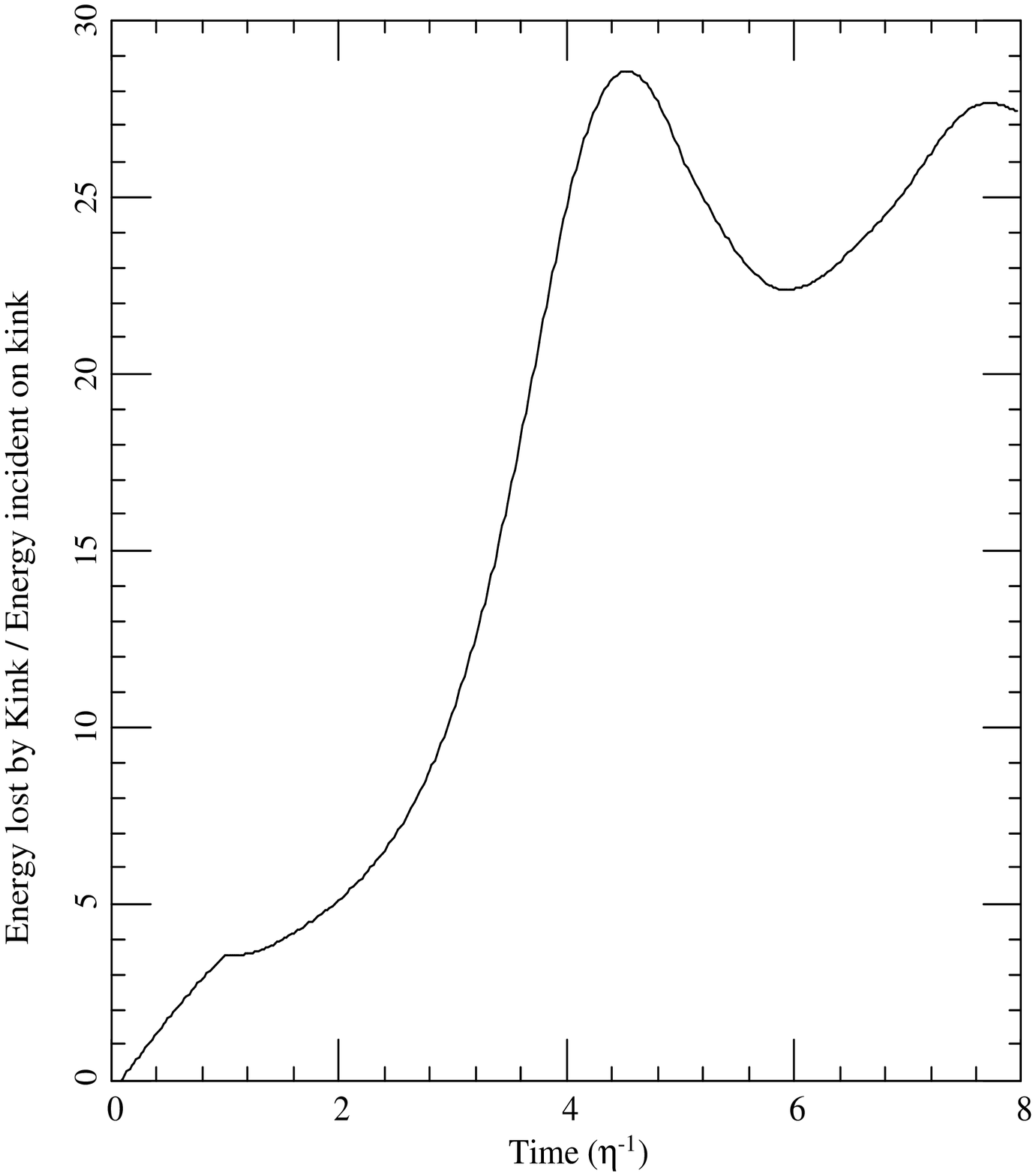}}
\bigskip
\caption{
Figure 1. Energy dissipation from kink vs time due to interaction with a
sinusoidal wave of `twist'}}\endinsert

In binary strings the presence of the twist modes thus appears to make
the kinks very 
unstable. Numerical simulations indicate that strings have a natural
tendency to intercommute and generate kinks \numer. 
Cosmologically, most of 
string length at any time is contained in one long string per
horizon. Numerous kinks should exist on the long strings from previous
self-intersections. In the case of binary strings the energy of the
kinks is continuously and rapidly drained into the twist modes, i.e the 
binary strings 
display an `elastic' resistance to developing sharp edges.
Consequently, the binary strings are likely to have more cusps
on them than ordinary strings. The reason for this is that kinks and
cusps are in some sense dual to each other \vach. The vectors ${
\partial  c^i \over \partial  \xi }$
and ${ \partial  d^i \over \partial  \xi }$ 
in \solconstr$\,$ describe curves on a unit sphere (the Kibble-Turok
sphere). Discontinuities in these curves represent kinks. But one
gets cusps if the curves intersect! 
For each kink smoothed out by winding
modes, a corresponding discontinuity in 
is removed; partial
smoothings reduce the jump in the discontinuity.
This causes the curves to intersect more,
generating more cusps.

The energy loss from cusps and kinks has been examined in the
literature \vach ,  \others. For non-superconducting strings, 
the loss of energy due to gravitational
radiation seems to be comparable in kinky and cuspy strings. For
superconducting strings, kinks can radiate large amounts of
power. Binary cosmic strings, even if superconducting, are however likely to
have no radiating kinks. 
For light
strings, where gravitational effects are small, particle radiation
\moreref$\,$
 through cusp annihilations can be
the dominant method for shrinkage of string loops \branden. In such a case,
long binary strings are expected to radiate copiously at all times. Recently
however, it has been argued that cusp annihilations may not be so
effective in particle production as thought previously \ruth. If this is
true then binary strings will radiate only by
gravitational processes. 
\vskip 2.0 cm
{\bf {Acknowledgements:}} The work of I.D was supported by the Department
of Energy under the grant DE-FG02-91ER40676.

\listrefs

\vfill\eject\end